\documentstyle[11pt,paspconf,epsf]{article}

\begin{document}

\title{Diffraction-limited 76~mas speckle-masking interferometry of
the carbon star IRC\,+10\,216 and related AGB objects with the SAO 6~m telescope}

\author{
G.\ Weigelt,
T.\ Bl\"ocker,
K.-H.\ Hofmann, and
R.\ Osterbart
}
\affil{
Max--Planck--Institut f\"ur Radioastronomie, Auf dem H\"ugel 69,
53121 Bonn, Germany}

\author{
Y.Y.\ Balega
}
\affil{
Special Astrophysical Observatory, Nizhnij Arkhyz,
Karachaevo--Cherkesia, 357147, Russia}

\author{
A.\ J.\ Fleischer and
J.\ M.\ Winters
}
\affil{
Technische Universit\"at Berlin, Institut f\"ur
Astronomie und Astrophysik, 10623 Berlin, Germany}

\begin{abstract}
We present high-resolution J-, H-, and K-band observations
of the carbon star IRC\,+10\,216.  The images were
reconstructed from 6~m telescope speckle interferograms using the
speckle masking bispectrum method.  The H image has the unprecedented
resolution of 70 mas.  The H and K images consist of at least five
dominant components within a 0.21 arcsec radius and a fainter
asymmetric nebula. The J--, H--, and K--band images seem to have an
{\sf X}-shaped bipolar structure.
A comparison of our images from 1995, 1996, 1997,
and 1998 shows that the separation of the two brightest components A
and B increased from $\sim$\,193 mas in 1995 to $\sim$\,246 mas in
1998.

The cometary shapes of component A in the H and J images and the
0.79~$\mu$m and 1.06~$\mu$m HST images suggest that the core
of A is not the central star, but the southern (nearer) lobe of the
bipolar structure.  The position of the central star is probably at or
near the position of component B, where the H--K color has its largest
value of H--K = 4.2.

If the star is located at or near B, then the components A, C, and
D are located close to the inner boundary of the dust shell at
separations of $\sim$\,200~mas $\sim$\,30~AU (projected distance)
$\sim$\,6~stellar radii for a distance of $\sim$\,150~pc, in agreement
with our 2-dimensional radiative transfer modelling.

In addition to IRC\,+10\,216 we studied the stellar disks and the dust
shells of several related objects.  Angular resolutions of~24 mas at
700~nm or 57~mas 1.6~$\mu$m were achieved. 


\end{abstract}

\section{Speckle masking observations of IRC +10 216}

IRC\,+10\,216 (CW Leo) is the nearest, brightest and best--studied
carbon star.  The central star of IRC\,+10\,216 is a long--period
variable with a period of about 650\,days.  Distance estimates between
110~pc and 170~pc were reported (see Crosas \& Menten 1997 and
references therein).  High--resolution IR observations of IRC +10
216 and of its circumstellar dust shell were reported by Dyck et al. 
(1991), Danchi et al.  (1994), McCarthy et al.  (1990), Christou et
al.  (1990), Osterbart et~al.  (1997), Weigelt et~al.  (1997, 1998),
and Haniff \& Buscher (1998).  Recent detailed radiative transfer
calculations for IRC\,+10\,216 were presented by Groenewegen (1997).

\begin{figure}
\centering
\mbox{\epsfysize=9cm\epsfbox{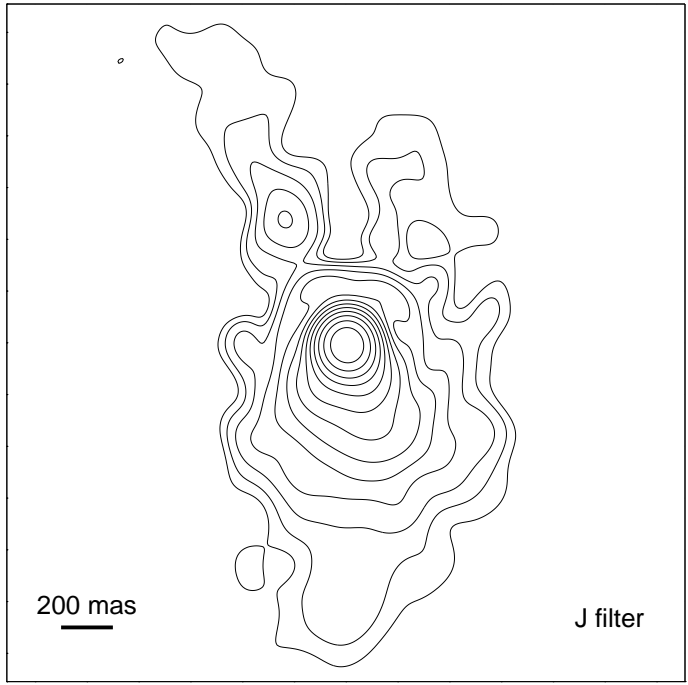}}
\caption{J--band image of IRC\,+10\,216 with 149~mas resolution. In all
images north is up and east to the left and the
contour spacing is 0.4 mag.}
\label{fig1}
\end{figure}

\begin{figure}
\centering
\mbox{\epsfysize=9cm\epsfbox{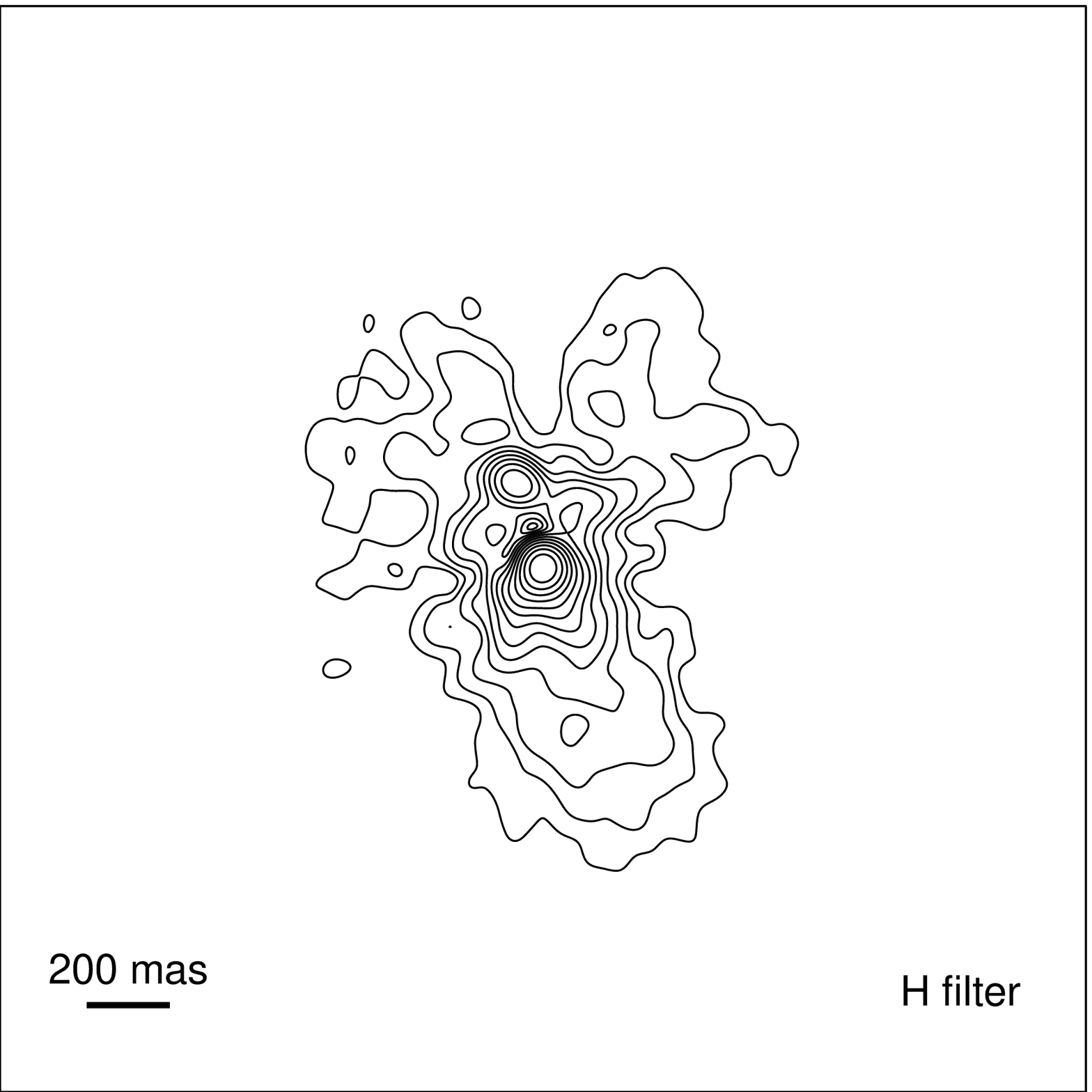}}
\caption{H--band image  of IRC\,+10\,216 with 70~mas resolution.} \label{fig2}
\end{figure}

\begin{figure}
\centering
\mbox{\epsfysize=9cm\epsfbox{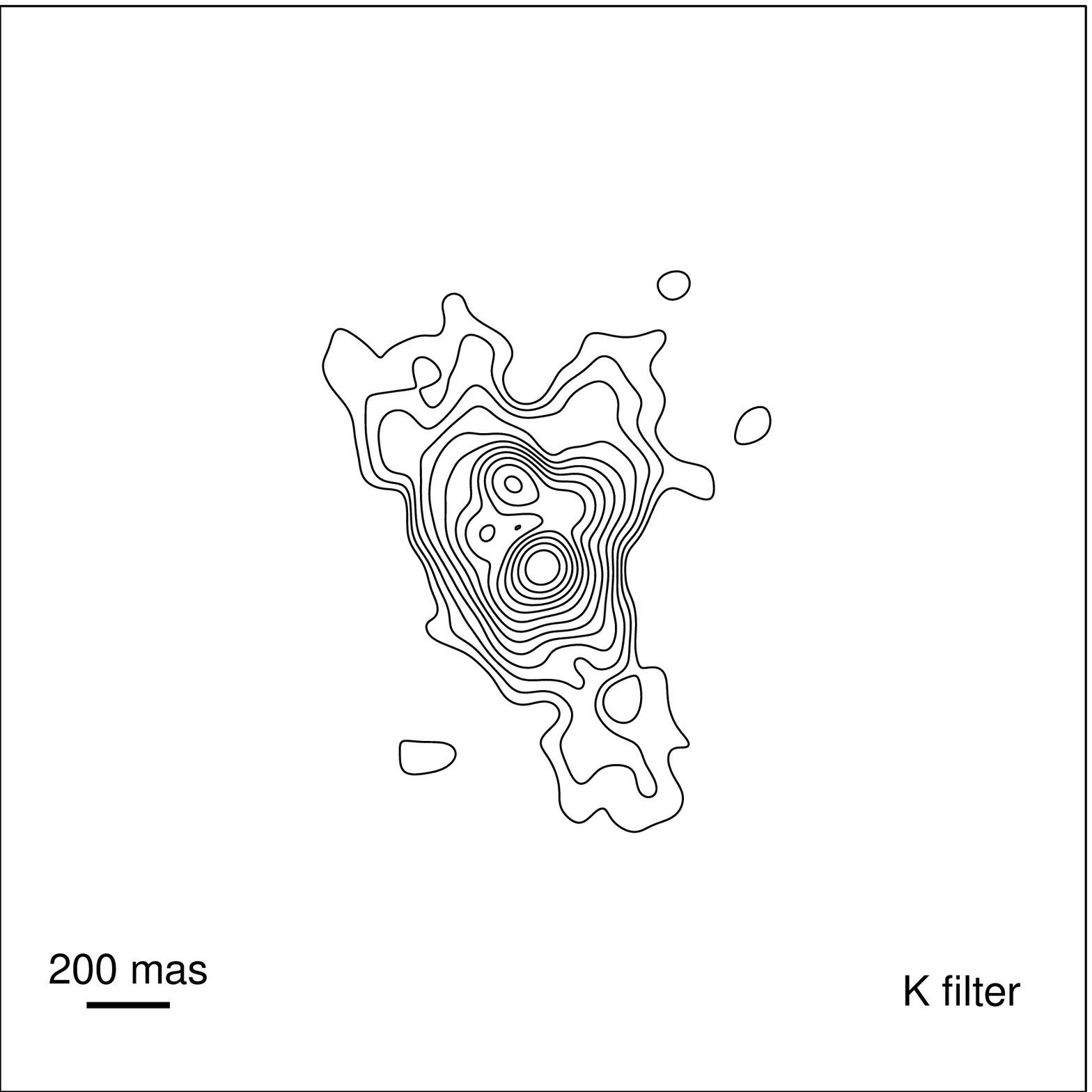}}
\caption{K--band image  of IRC\,+10\,216 with 87~mas resolution.} \label{fig3}
\end{figure}

The IRC\,+10\,216 speckle interferograms were obtained with the 6\,m
telescope at the Special Astrophysical Observatory in Russia and our
NICMOS 3 camera at four different epochs.  The high--resolution J, H,
and K images (Figs.~\ref{fig1} to \ref{fig3}) were reconstructed from
the speckle interferograms using the speckle masking bispectrum method
(Weigelt 1977, Lohmann et al.\ 1983, Weigelt 1991).  The resolution of
the J (April 2, 1996), H (January 23, 1997), and K (January 23, 1997)
images are 149~mas, 70~mas, and 87~mas, respectively.  We denote the
resolved components in the H and K image as A, B, C, and D (in the
order of decreasing peak intensity; Weigelt et~al.\ 1998).

The separation of the components A and B was measured for four
different epochs.  The four obtained separations are:  193~mas
(October~1995, phase 0.88), 201~mas (April~1996, phase 0.15), 212~mas
(January~1997, phase 0.62), and 246~mas (June~98, phase 0.39).  The
linear regression fit gives a velocity for this motion of
$\sim$\,20~mas/yr.  Assuming a distance of 150~pc (Crosas \& Menten
1997) for IRC\,+10\,216, this tangential velocity transforms to 14~km/s. 
This velocity is smaller than the value of the radial expansion
velocity of 15~km/s of the circumstellar matter (see Gensheimer \&
Snyder 1997 and references therein).  From the present data there is
no evidence that the motion inside the nebula is correlated with the
stellar pulsation cycle.

\section{Discussion}

{\it The core structure and the faint nebula in the H and K images.}
The multi--component structure of the IRC\,+10\,216 K--band image (of
the innermost \linebreak
300~mas~$\times$~300~mas) has already been reported by
Weigelt et~al.\ (1998) and Haniff \& Buscher (1998).  This bright
inner region is surrounded by a larger faint bipolar nebula (with $\sim$\,1\%
of the peak brightness of A).  On the northern side of this faint
nebula two arms of the nebula can be seen at position angles of
roughly 30\deg  (NE) and 340\deg  (NW) with respect to component
A.  On the southern side one arm is present at about 210\deg (SW). 
At 160\deg (counterside of the NW-arm) the nebula is much fainter. 
It may, therefore, be adequate to take the direction from component A
to component B (about 20\deg{}  position angle) as the direction of
the main axis (see also Kastner \& Weintraub 1994).

{\it The bipolar J--band image.} The J--band image (Fig.~\ref{fig1}) and the
0.79~$\mu$m and 1.06~$\mu$m HST images (Haniff \& Buscher 1998) show a
bipolar structure of the nebula.  The southern lobe has a cometary
shape whereas the northern region of the images shows two arms
reminiscent to (but much weaker than) the {\sf X}--shaped
structure of the Red Rectangle (see Men'shchikov et al.\ 1998).

{\it H-K color image.} In a square aperture of 1.6~arcsec the H and K
magnitudes were determined to be K=2.5 and H=5.7.  The integral
color in this field is thus H-K=$3.2\pm0.2$.  Our high--resolution
H-K color image (Osterbart et al.\ 1999) shows that the components B,
C, and D are rather red (H-K$\approx$4.2) in comparison with the
integral color.  The brightest component A (H-K=3.2) as well as the
{\em cometary} southern tails (H-K=2 to 3) in the H and J image are
bluer.  These tails are a striking feature best seen in the
high--resolution H--band image (Fig.~\ref{fig2}).  This structure and its
relatively blue color suggest that component A is produced to a significant
fraction by scattering of stellar light.  This morphology
can easily be understood under the assumption that the density of the
dust shell is not sperically symmetric.  The polar axis probably
points with its southern side towards us.  Component B may then be at
or near the position of the star, strongly obscured and reddened by
its environment.  This suggests that the main axis of the nebula has a
position angle of $\sim$\,20\deg.

{\it Relative motion of A and B.} The relative motion of the nebula
components is clearly not related to the stellar variablity which has
a period of about 650~days.  It may thus be related to either an
overall expansion or a variability of the dust shell with a period
significantly larger than the stellar pulsation period (cf.  Winters
et~al.\ 1995).

{\it Bipolar structure, clumpiness, and position of the central star.}
Theoretical models treating the dust formation mechanism in the
envelope of long--period variable carbon stars (Winters et~al.\
1994a,b) predict that dust formation does not occur permanently but
ceases regularly.  Periods of this mechanism may be significantly
different from the stellar pulsation period.  This leads us to the
conclusion that in our IRC\,+10\,216 images we see the inner boundary of
the dust shell moving outwards with approximately 15~km/s.  The
components in the central region of IRC\,+10\,216 have separations of
the order of only a few stellar radii and the travel time of newly
formed dust over these distances is only a few years.  The components
A to D may be caused, for instance, by large convection cells in the
stellar atmosphere leading to asymmetric and stochastic fluctuations
of the dust formation and thus to disturbances of the overall bipolar
geometry.

Consistent with preliminary results from our radiative transfer
calculations (Men'shchikov et~al.\ 1999, in preparation), the cometary
shapes of A in the H and the J images and the 0.79~$\mu$m and
1.06~$\mu$m HST images (Haniff \& Buscher 1998) suggest that the core of A
is {\em not} the central star, but the central star is at or near B,
between the northern and southern J--band lobes separated by
$\sim$\,500~mas.  B is very red (H-K$\approx$4.2) and, therefore, it is
not visible in the J--band image.  If the star is now at or near B,
then the components A, C, D and the northern components in the J image
are probably located at the inner boundary of the dust shell.  The
radiative transfer models imply that the polar axis is inclined by
intermediate angles (roughly 50\deg) with respect to the plane of the
sky.

\section{Diffraction-limited speckle masking studies of R Cas, AFGL
2290, and the Red Rectangle}

In addition to IRC +10 216 we have studied the wavelength dependence
of the size and the shape of the Mira stars R Cas (Weigelt et al.\
1996) and R Leo, the aspherical dust shell of the oxygen-rich AGB star
AFGL 2290 (Gauger et al.\ 1999), and the bipolar structure of the Red
Rectangle (Men'shchikov et al.\ 1998).  One-dimensional radiative
transfer modelling of AFGL 2290 and two-dimensional radiative
transfer modelling of the Red Rectangle was used for the
interpretation of the observations.

\acknowledgments
This research has made use of the SIMBAD database, operated at CDS,
Strasbourg, France.

\end{document}